\documentclass[prd,onecolumn,showpacs,nofootinbib,preprintnumbers,superscriptaddress]{revtex4}
\usepackage{amsmath}\usepackage{amsfonts}\usepackage{graphicx}\usepackage{hyperref}\usepackage{mathbbol}
\def\be{\begin{equation}}   \def\ee{\end{equation}}   \def\bea{\begin{eqnarray}}    \def\eea{\end{eqnarray}}  \def\no{\nonumber}
    \def\d{{\rm d}}         \def\k{\kappa}       
\def\r{\right}            \def\l{\left}
\def\f{\frac}    \def\l{\left}   \def\r{\right}

\DeclareMathOperator{\arcsec}{arcsec}

\DeclareMathOperator{\arcsinh}{arcsinh}
\DeclareMathOperator{\arctanh}{arctanh}

\DeclareMathOperator{\arccoth}{arcCoth}

\begin{document}
\title{Non-linear Schr\"{o}dinger-type formulation of scalar field cosmology: two barotropic fluids and exact solutions}
\date{\today}


\author{Chonticha Kritpetch}\email{chontichakr57@email.nu.ac.th}
 \affiliation{The Institute for Fundamental Study ``The Tah Poe Academia Institute", Naresuan University, Phitsanulok 65000, Thailand}
 \author{Jarunee Sanongkhun}\email{jaruneen57@email.nu.ac.th}
 \affiliation{The Institute for Fundamental Study ``The Tah Poe Academia Institute", Naresuan University, Phitsanulok 65000, Thailand}
\author{Pichet Vanichchapongjaroen}\email{pichetv@nu.ac.th}
 \affiliation{The Institute for Fundamental Study ``The Tah Poe Academia Institute", Naresuan University, Phitsanulok 65000, Thailand}
\author{Burin Gumjudpai}\email{Corresponding: buring@nu.ac.th}
 \affiliation{The Institute for Fundamental Study ``The Tah Poe Academia Institute", Naresuan University, Phitsanulok 65000, Thailand}
\affiliation{Thailand Center of Excellence in Physics, Ministry of Education, Bangkok 10400, Thailand}


\begin{abstract}
Time-independent non-linear Schr\"{o}dinger-type (NLS) formulation of FRW cosmology with canonical scalar field are considered in case of two barotropic fluids. We derived Friedmann formulation variables in terms of NLS variables. Seven exact solutions found by D'Ambroise \cite{DAmbroise:2010dgl} and one new found solution are explored and tested in cosmology. The result suggests that time-independent NLS formulation of cosmology case should be upgraded to the time-dependent case.
\end{abstract}

\pacs{98.80.Cq}

\date{\today}

\vskip 1pc

\maketitle \vskip 1pc
\section{Introduction}
Present universe is under acceleration phase \cite{Masi:2002hp} and the cosmic picture must be consequence of inflationary expansion at very early time \cite{inflation}.
The accelerating expansion could result from dynamical canonical or phantom scalar field with
time-dependent equation of state $w_{\phi}(t) < -1/3$
or from modification of general
relativity (see e.g. \cite{Nojiri:2006ri, Padmanabhan:2004av}). Conventional FLRW cosmology is the Einstein field equations, i.e. the Friedmann and acceleration equations with conservation in form of the fluid equation. The system is sourced by canonical (or phantom) scalar field and barotropic perfect fluids resulting in the cosmic kinematics. There is an alternative mathematical approach to the same system in which the cosmological equations are expressed in form non-linear Schr\"{o}dinger (NLS) equation. We review this in the following.

 Ermakov system \cite{Er1, Er2}, which is a pair of non-linear second-order ordinary differential equations, was noticed to have a connection to standard FLRW cosmology sourced by a barotropic perfect fluid and a self-interacting canonical scalar field minimally coupled to gravity. This provides alternative analytical approach to the cosmological system \cite{Hawkins:2001zx}. One-dimensional Ermakov system decouples to single equation dubbed the Ermakov-Pinney (or Milne-Pinney) equation \cite{Er1, Pinney, Milne},
\be
\ddot{b}  + Q(t) b  \;=\; \f{\lambda}{b^3}\,,    \label{EP}
\ee
where $b = b(t) \equiv u^{-1}(t) = a^{n/2}$. Function $a$ is the scale factor and $t$ is cosmic time. Dot is $\d /\d t$. Albeit its non-linearity, its general solution is a superposition of particular solutions of a related linear second-order ordinary differential equation when the constant $\lambda = 0$ \cite{Pinney, lut}. As discussed in \cite{Hawkins:2001zx}, $Q(t)$ and $\lambda$ reads\footnote{Here we change the expression of variables in \cite{Hawkins:2001zx} so that it matches the later literatures.}
\be
Q(t) = \f{\kappa^2 n}{4} {\dot{\phi}}^2 \;\;\;\; \text{and}   \;\;\;\; \lambda =  -\f{D n^2 \kappa^2}{12}.
\ee
The system above is related to FLRW cosmology of the flat ($k=0$) case of the system,
\bea
H^2 & = &  \f{\kappa^2}{3}\l(\rho_{\phi}  +  \f{D}{a^n} \r) -  \f{k}{a^2}, \\
\epsilon(\ddot{\phi} + 3 H \dot{\phi})& = & - \f{\d V}{\d \phi}.
\eea
where the speed of light $c\equiv 1$, $\kappa^2 \equiv 8 \pi G$, $D \geq 0$ is proportional constant, $\epsilon = 1$ or $-1$ for canonical or phantom field cases. The scalar field density is, $\rho_{\phi} = (1/2)\epsilon \dot{\phi}^2 + V(\phi)$, the scalar field pressure is,  $p_{\phi} = (1/2)\epsilon \dot{\phi}^2 - V(\phi)$. Barotropic fluid pressure and density are, $p_{\gamma} = w_{\gamma}\rho_{\gamma}$ and $\rho_{\gamma} = {D}/{a^n}$ where $n = 3(1+w_{\gamma})$. With further reparametrization $x(t) = \int u \, \d t$, the Ermakov-Pinney equation (\ref{EP}) is expressed as time-independent one-dimensional linear Schr\"{o}dinger equation,
\be
u'' + \l[E - P(x) \r] u(x)  = 0,
\ee
where $' \equiv \d/\d x $,  $\: E = -(\kappa^2 n^2 D)/12 $ and $P(x) = (\kappa^2 n/4)\epsilon (\d \phi/\d x)^2 $.   Hence flat FLRW cosmology with scalar field and a barotropic fluid can be described by a linear Schr\"{o}dinger equation. This relation is also applicable in case of RSII braneworld \cite{Hawkins:2001zx}. The connection between FLRW scalar field cosmologies to non-linear partial differential equations such as the Ermakov-Penny equation in 2+1 dimensions \cite{WK2_1} and 3+1 dimensions were further studied and blowup solutions are found, giving hope to have relevance to non-linear quantum cosmology \cite{WK3_1}. Non-flat ($k \neq 0$) case extension of the FLRW system is reported in \cite{WKCPG} and Bianchi I and V extension of the approach are also made. It is also found that Bianchi I Einstein field equation with scalar field and a perfect fluid is equivalent to linear Schr\"{o}dinger equation \cite{D'Ambroise:2007gm}.
  Cosmology in form of Ermakov-Penny equation with $k>0$ is found to be corresponding to two-dimensional Bose-Einstein condensates \cite{Lidsey:2003ze}. Perturbative scheme of the solution of the Ermakov-Pinney equation was developed in connection to generalized WKB method  \cite{Kamenshchik:2005kf}.
The work by \cite{Williams:2005bp} shows that a generalized Ermakov-Milne-Pinney (EMP) equation is completely equivalent to the FLRW scalar field cosmology (including the non-flat case). It comfirms and generalizes the result of \cite{Hawkins:2001zx}. The generalized EMP equation later was found to be equivalent to the NLS equation,
 \bea u''(x) + \left[E-P(x)\right]
u(x)
 = -\frac{nk}{2}u(x)^{(4-n)/n}\,,   \label{schroeq} \eea
 providing alternative approach to the FLRW scalar field cosmology with quantum-mechanical formulation \cite{D'Ambroise:2006kg}.

In the NLS-Friedmann correspondence, inputs are assumed scale factors which enable us to obtain exact solutions for a non-flat Friedmann universe with a barotropic fluid and a scalar field  \cite{Gumjudpai:2007bx}.
Recently, parametric solutions of non-linear ordinary differential equation of which the special cases are homogeneous and inhomogenous cosmologies and Bose-Einstein condensation correspondence, are found \cite{DW2013}.
These literatures motivated studies on the NLS formulation of scalar field cosmology assuming scale factors functions  \cite{Gumjudpai:2007qq, Phetnora:2008mf, Gumjudpai:2009ws, Gumjudpai:2008mg}. Detail of the NLS formulation is presented in D'Ambroise's dissertation \cite{DAmbroise:2010dgl} which also gives larger classes of solution of the system.  Further connection in case of time-dependent NLS equation and Friedmann scalar field cosmology was studied and it is possible to fulfill the need of non-perturbative quantum description of gravity and cosmology  since it establishes correspondence between quantum and gravitational systems \cite{Lidsey:2013osa}.

Here in this work, we investigate the time-independent NLS equation in connection to Friedmann formulation in the case of two barotropic fluids with a canonical scalar field. We consider and analyse solutions of the NLS system of the two-fluid case based on possible $u(x)$ solutions reported in \cite{DAmbroise:2010dgl}. We try to interpret the given possible solutions.

\section{Equation of motion}
Considering a FRW universe sourced by two non-interacting perfect fluids
and a minimally coupled scalar field $\phi$ with potential $V(\phi)$,
density and pressure of the fluids are given by
\be
\rho_1 = \frac{D_1}{a^n},\qquad
\rho_2 = \frac{D_2}{a^m},     \label{densityev}    \ee
\be
p_1 = \l( \frac{n-3}{3} \r) \frac{D_1}{a^n},\qquad
p_2 = \l( \frac{m-3}{3} \r) \frac{D_2}{a^m},
\ee
whereas the scalar field density and pressure are given by
$
\rho_\phi = (1/2)\epsilon\dot\phi^2+V(\phi),\;
p_\phi = (1/2)\epsilon\dot\phi^2-V(\phi)
$ as above. The scalar equation of state is $w_{\phi} = p_{\phi}/\rho_{\phi}$.
The dynamics are governed by the Friedmann equation,
\be
H^2=\frac{\kappa^2}{3}\rho_{\text{tot}}-\frac{k}{a^2}=
\frac{\kappa^2}{3}\left[\frac12\epsilon\dot\phi^2+V+\frac{D_1}{a^n}+\frac{D_2}{a^m}\right]-\frac{k}{a^2},
\ee
and by acceleration equation,
\bea
\frac{\ddot a}{a} &=& -\frac{\kappa^2}{6}(\rho_{\text{tot}}+3p_{\text{tot}}),  \\ &=& -\frac{\kappa^2}{6}\left[2\epsilon\dot\phi^2-2V+(n-2)\frac{D_1}{a^n}+(m-2)\frac{D_2}{a^m}\right].
\eea
Note that the Klein-Gordon equation is a consequence of the above two equations.
It is sufficient to consider only the Friedmann equation and acceleration equation. Therefore
we have
\bea
\label{kin}
\epsilon\dot\phi(t)^2 &=& -\frac{2}{\kappa^2}\left[\dot H-\frac{k}{a^2}\right]-\frac{n D_1}{3a^n}-\frac{mD_2}{3a^m},\nonumber\\
\label{pot}
V(\phi) &=& \frac{3}{\kappa^2}\left[H^2+\frac{\dot H}{3}+\frac{2k}{3a^2}\right]+\l(\frac{n-6}{6}\r)\frac{D_1}{a^n}  + \left(\frac{m-6}{6}\right)\frac{D_2}{a^m}.
\eea
In general, once we specify $a(t),D_1, D_2,n,m,k,$ we can immediately
obtain $\epsilon\dot\phi(t)^2$ and $V(\phi).$ The value for $n$ or $m$ implies types of barotropic fluids, for instance, $n = 0$ for $w_{\gamma} = -1$, $n = 2$ for $w_{\gamma} = -1/3$,   $n = 3$ for $w_{\gamma} = 0$ (dust), $n = 4$ for $w_{\gamma} = 1/3$ (radiation), $n = 6$ for $w_{\gamma} = 1$ (stiff fluid).

\section{NLS Formulation}
In order to connect the Friedmann formulation to the NLS formulation, we define\footnote{We add $D_2$ contribution to $P(x)$ rather than adding to $E$ because $E$ must be constant in according to the solutions listed in table \ref{tab:7solutions}},
\bea
u(x) &\equiv & a(t)^{-n/2}, \qquad
E\equiv - \frac{\kappa^2n^2}{12}D_1,   \label{ua}  \\
P(x) & \equiv & \frac{\kappa^2n}{4}a(t)^n\epsilon\dot\phi(t)^2+\frac{m D_2}{12}\kappa^2 n a^{n-m},  \label{PxEq}
\eea
where $\dot x(t)=u(x).$ The equation (\ref{kin}) then becomes a non-linear Schr\"odinger equation
\be\label{NLS}
u''(x)+\left[E-P(x)\right]u(x) = -\frac{nk}{2}u(x)^{(4-n)/n}.
\ee
We can express $\epsilon\dot\phi(t)^2, V(\phi)$ and the other cosmological quantities
as
\be
\epsilon\dot\phi^2 = \frac{4}{\kappa^2n}uu''+\frac{2k}{\kappa^2}u^{4/n} + \frac{4E}{\kappa^2n}u^2-\frac{mD_2}{3}u^{2m/n},
\ee
\be
V = \frac{12}{\kappa^2n^2}(u')^2-\frac{2P}{\kappa^2n}u^2+\frac{12E}{\kappa^2n^2}u^2
+\frac{3k}{\kappa^2}u^{4/n}+\left(\frac{m-6}{6}\right)D_2u^{2m/n},
\label{Vaa}  \ee
\bea
\rho_\phi &=& \frac{12}{\kappa^2n^2}(u')^2+\frac{12E}{\kappa^2n^2}u^2+\frac{3k}{\kappa^2}u^{4/n}-D_2u^{2m/n},  \no \\
&=&   \f{12}{\kappa^2n^2}(u')^2  - u^2D_1 +  \f{3k}{\kappa^2}u^{4/n} - D_2u^{2m/n}.
\eea
\bea
p_\phi (= \rho_\phi-2V) &=& -\frac{12}{\kappa^2n^2}(u')^2 + \frac{4P}{\kappa^2n}u^2- \frac{12E}{\kappa^2n^2}u^2-\frac{3k}{\kappa^2}u^{4/n}-\left(\frac{m-3}{3}\right)D_2u^{2m/n},   \no \\
      & = &   -\frac{12}{\kappa^2n^2}u'^2  +  \frac{4}{\kappa^2n}uu'' -
       \frac{k}{\kappa^2}u^{4/n} - \left(\frac{n-3}{3}\right) u^2D_1
                       -\left(\frac{m-3}{3}\right)D_2u^{2m/n},  \\
\eea
\be
\rho_{\text{tot}} = \frac{12}{\kappa^2n^2}(u')^2+\frac{3k}{\kappa^2}u^{4/n}-D_2u^{2m/n} \left(=\frac{3}{\kappa^2}\left[H^2+\frac{k}{a^2}\right]\right),
\ee
\be
p_{\text{tot}} \left(=-\frac{2}{\kappa^2}\left[\dot H+H^2+\frac{\kappa^2}{6}\rho_{\text{tot}}\right]\right) = -\frac{12}{\kappa^2n^2}(u')^2 +\frac{4}{\kappa^2n}uu''-\frac{k}{\kappa^2}u^{4/n},
\ee
\be
H=-\frac{2}{n}u',\qquad
\dot{H} = -\frac{2}{n}uu'',
\ee
\be
\ddot\phi=\pm\frac{P'u^2+2uu'\left(P-\frac{m^2D_2\kappa^2u'u^{2(m-n)/n}}{12}\right)}{\kappa\sqrt{n\epsilon}\sqrt{P-\frac{D_2mnu^{2(m-n)/n}\kappa^2}{12}}},\qquad
3H\dot\phi = \mp\frac{12 u' u\sqrt{P-\frac{1}{12}D_2 \kappa ^2 m n u^{\frac{2 (m-n)}{n}}}}{n\kappa  \sqrt{\epsilon n}}\,.
\ee
Using these relations, we recover the NLS equation (\ref{NLS}) with the NLS potential,
\be
P(x)  =   \f{u''}{u} + \f{kn}{2} u^{(4/n)-2}  + E \no
\ee
where NLS kinetic energy is $T = - ({u''}/{u}) - ({kn}/{2}) u^{(4/n)-2} $. Note that $u' = \dot{u}/u$  and  $u''  =  u^{-1} (\d u'/\d t)$.
If expressed in term of density parameters
\bea
\Omega_1 \equiv \frac{\rho_1}{\rho_{\rm c}} = \frac{n^2D_1\kappa^2u^2}{12(u')^2}, \;\;\;\;\;\;\;
\Omega_2 \equiv \frac{\rho_2}{\rho_{\rm c}} = \frac{n^2D_2\kappa^2u^{2m/n}}{12(u')^2}, \;\;\;\;\;\;\;
\Omega_k \equiv \frac{\rho_k}{\rho_{\rm c}} = -\frac{k}{a^2H^2}  = -\frac{kn^2}{4(u')^2u^{-4/n}},
\eea
where
\be
\rho_{\rm c} \equiv \rho_{\rm tot}-\frac{3k}{\kappa^2 a^2} = \frac{3H^2}{\kappa^2} = \f{12 u'^2}{\k^2 n^2}    ,\;\;\;\;\;\;\;\;
\rho_k \equiv -\frac{3k}{\kappa^2 a^2} = -\f{3 k u^{4/n}}{\k^2},
\ee
such that the Friedmann equation $
\Omega_\phi \equiv {\rho_\phi}/{\rho_{\rm c}} = 1-\Omega_1-\Omega_2-\Omega_k
$
 is,
 \be
 \Omega_\phi  =  1  - \f{n^2 \k^2}{12 u'^2}  \l(  D_1 u^2  + D_2 u^{2m/n} -  \f{3 k u^{4/n} }{\k^2}      \r).
 \ee
Here we consider only non-phantom case, i.e. $\epsilon = 1$.

\section{NLS exact solutions}

Following the D'Ambroise thesis \cite{DAmbroise:2010dgl}, we consider the NLS equation,
\be
u''(\sigma)+[E-P(\sigma)]u(\sigma) = \dfrac{F}{u(\sigma)^{C}},    \label{NLSGenn}
\ee
 where $E,F$ and $C$ are constants and
 \bea
 D_1 = -\f{12 E}{n^2 \kappa^2}, \;\;\;\;  F = -\f{n k}{2},  \;\;\;\;  C = \f{n-4}{n}.
 \eea

D'Ambroise demonstrates that there are at least seven exact solutions of NLS for single barotropic-fluid case \cite{DAmbroise:2010dgl}. Here we apply the solutions to  the NLS equation for two barotropic fluids. Contribution of the second fluid is expressed as an additional term in $P(x)$ as seen in the equation (\ref{PxEq}).  We quote table of solutions from table E.1 of the previous work \cite{DAmbroise:2010dgl} into Table \ref{tab:7solutions} of this work where minor notation here is altered from \cite{DAmbroise:2010dgl}, i.e. $\sigma \rightarrow x $, $a_0 \rightarrow e_0$ and $\theta \equiv 1$. Features of the NLS formulation are the benefits of having an alternative way of solving for (1) scalar field exact solutions (as in \cite{Gumjudpai:2007bx}) and (2) scale factor solutions. Here we emphasize our studies on the scale factor solutions.

\begin{table}[h!]
	\caption {The NLS exact solutions given by  J.~D'Ambroise \cite{DAmbroise:2010dgl}} \label{tab:7solutions}
	\begin{tabular}{|l|l|l|l|l|l|}
		\hline
		& Solutions: $u(x)$                                              & $P(x)$                                               & $E$ & $F$     & $C$  \\ \hline
		$1$ & $e_0x^2+b_0x+c_0$               & $(2e_0+d_0)/(e_0x^2+b_0x+c_0)$ & 0 & $-d_0$ & $0$  \\ \hline
		$2$ & $e_0\cos^2(b_0x)$ &   $2b_0^2\tan^2(b_0x)$                                                 & $2b_0^2$  & $0$      &  arbitrary \\
		& & $4b_0^2\tan^2(b_0x)$& $0$ & $-2b_0^2e_0$& $0  $\\ \hline
		3 &$e_0\tanh(b_0x)$                                                   &$c_0$ & $c_0+2b_0^2$  & $2b_0^2/e_0$      &  $-3$   \\ \hline
		$4$& $e_0e^{(-x\sqrt{-c_0})}-b_0e^{x\sqrt{-c_0}}$              & $0$  &   $c_0<0$    &   $0$ & arbitrary \\ \hline
		$5$ &  $(e_0/x)e^{c_0x^2/2} $                                                 &    $c_0^2x^2+2/x^2+b_0$                                                & $c_0+b_0$  &    $0$ & arbitrary   \\ \hline
		$6$ &  $-e_0\cosh^2(b_0x)$                                                 & $2b_0^2\tanh^2(b_0x)+c_0  $                                                 & $c_0-2b_0^2$   &        $0$ & arbitrary \\ \hline
		$7$ & $e_0/x^{b_0}  $                                                 &    $\frac{b_0(b_0+1)}{x^2}+c_0$                                                &  $c_0$  &        $0$ & arbitrary \\ \hline
	\end{tabular}
\end{table}

\subsection{Solution 1}
The first solution of equation (\ref{NLSGenn}) is
\be
u(x) = \dot{x} = e_0 x^2 + b_0 x + c_0\,,   \label{line1sol}
\ee
where $E = 0, F = -d_0$ and $C = 0$. These imply
$D_1 = 0, n = 4$ and $ k = d_0/2 $ and equation (\ref{NLSGenn}) in this case is
\be
u''(x) - P(x)u(x) = -d_0.
\ee
Hence $D_1$ represents the radiation fluid since $n = 4$ (see equation (\ref{densityev})). However there is no radiation density for this solution since $D_1 = 0$, hence there are only fluid $D_2$ and curvature $k = d_0/2$.
\begin{itemize}
\item {\bf Case 1.1:} $e_0 \neq 0$ \\
The solution is reported in \cite{DAmbroise:2010dgl} as,
\be
x(t)  =  \f{1}{2 e_0} \l\{ \sqrt{-\Delta} \tan\l[ \f{\sqrt{-\Delta}}{2} (t - t_0) \r] - b_0 \r\}\,,
\ee
where $\Delta = b_0^2 - 4 e_0 c_0 < 0$ and
\be
u(t)  =  -\f{\Delta}{4 e_0}\sec^2 \l[\f{\sqrt{-\Delta}}{2}(t-t_0)  \r]\,.
\ee
The coefficients $e_0, c_0$ must take the same signs, i.e. $e_0 > 0$ when $c_0 > 0$ or $e_0 < 0$ when $c_0 < 0$ so that the condition $\Delta < 0$ is satisfied.  From $u = a^{-n/2}$ in equation (\ref{ua}) hence the scale factor is
\be
a(t)  =   \l\{ - \f{4 e_0}{\Delta} \cos^2 \l[ \f{\sqrt{-\Delta}}{2} (t - t_0) \r]  \r\}^{2/n}.
\ee
In form of redshift, $1+z = a(t_0)/a(t)$ hence
\be
z(t)  =  \l\{ \sec^2\l[\f{\sqrt{-\Delta}}{2}(t-t_0)   \r] \r\}^{2/n} - 1\,, \;\;\;\;{\text{and}}\;\;\;\;
t -t_0    =   \f{2}{\sqrt{-\Delta}} \l\{\arcsec[(z+1)^{n/4}] \r\}\,.
\ee
The Hubble rate is derived,
\be
H(t)   =  -\f{2\sqrt{-\Delta}}{n} \tan \l[ \f{\sqrt{-\Delta}}{2}(t-t_0) \r] \,, \;\;\;\;{\text{or}}\;\;\;\;
H(z)  =   -\f{2\sqrt{-\Delta}}{n} \tan \l\{  \arcsec[(z+1)^{n/4}]        \r\}\,.
\ee
For $t > t_0$, Hubble rate is negative, the universe contracts and for $t < t_0$ the universe expands. Both cases blow up at some finite values of the tan function.

\item {\bf Case 1.2:} $e_0 = 0$ \\
The wave function reduces to $ u(x) = b_0 x + c_0$ and the solution is\footnote{Here we give correction to the result in \cite{DAmbroise:2010dgl}.}
\be
x(t)  =  \f{1}{b_0}\l[e^{b_0(t-t_0)} - c_0\r]\,,
\ee
and for $b_0 \neq 0 $,
\be
u(t)  =    e^{b_0(t-t_0)}\,.
\ee
The scale factor is hence
\be
a(t)  =     e^{-2 b_0(t-t_0)/n}\,,
\ee
hence
\be
z(t) = e^{2 b_0 (t-t_0)/n} - 1\,,
\ee
and $H = -2b_0/n = H_0$ is a constant Hubble rate. Since $n = 4$ hence $b_0 = - 2 H_0$. This could give either positive or negative constant $H_0$ depending on the sign of $b_0$. For negative $b_0$, the expansion is of the de Sitter type.
\end{itemize}
Although, we have solutions for the cases 1.1 and 1.2, it does not make sense to have zero density of the first fluid, $D_1 = 0$
but having $n=4$. Fluid density with zero value must remain zero forever. The appearing of $n = 4$ in expressions of density and pressure makes no sense.



\subsection{Solution 2}
The second solution expresses that \be u(x) = e_0\cos^2(b_0x)\,. \label{eqsol2}  \ee The conditions satisfying this solution are
\begin{itemize}
	\item  {\bf Case 2.1}: $ E=2b_0^2$, $F = 0$ and $C$ is arbitrary. The form of $E$ gives $D_1 = -24 b_0^2/ n^2 \kappa^2 < 0$, i.e. negative density. The condition $ F = 0 =  -n k/2$ is considered into three subcases. First,  $k=0$ and arbitrary $n$ give arbitrary value of $C$ and $D_1 < 0$ for $b_0 \neq 0$. Secondly, $n=0$ and arbitrary $k$ correspond to $C = \infty$ and $D_1 = \infty$.
Thirdly, $k=0$ and $n=0$ imply $C = \infty$ and $D_1 = \infty$. Having negative or infinity values of density proportional constant (of the barotropic fluid) is nonphysical and is not of our interest.
\item  {\bf Case 2.2}: $ E=0$, $F = -2b_0^2 e_0$ and $C=0$. This gives $D_1 = 0, n=4$ (radiation) and $k = b_0^2 e_0$. There is no radiation density in this solution although we know that $n$ must be of the radiation. Hence the system is of the universe with arbitrary $k= b_0^2 e_0$ and a second fluid with $m$ value of barotropic equation of state with density $D_2$.
\end{itemize}
The solution equation (\ref{eqsol2}) corresponds to
\be
x(t) = \frac{1}{b_0}\arctan\l[e_0b_0(t-t_0)\r]\,, \;\;\;\;\;\text{and}\;\;\;\;\;
u(t) = \dot{x}(t) = \frac{e_0}{1+e_0^2b_0^2(t-t_0)^2}.
\ee
The scale factor solution is found as
\be
a(t) = \l[\frac{1+e_0^2b_0^2(t-t_0)^2}{e_0}\r]^{2/n}\,,
\ee
where $e_0 \neq 0$. As $n = 4$ hence
\be
H(t) = \dot{a}/{a} = \frac{e_0^2b_0^2(t-t_0)}{e_0^2b_0^2(t-t_0)^2 + 1},
\ee
and time-redshift relation is
\be
z(t)  = \sqrt{\frac{1}{e_0^2 b_0^2 (t-t_0)^2+1}} - 1\,.
\ee
We hence write
\be
H(z) =  e_0b_0(z+1)\sqrt{-z(z+2)}\,.
\ee
The valid range of redshift is $z \in (-2,0)$ which is not realistic.  Negative density, $D_1 < 0$ of the case 2.1 is not physical.
The case 2.2 has the same problem of the case 1.1 and 1.2 such that $D_1 = 0$.

\subsection{Solution 3}
The given solution is
\be u(x) = e_0\tanh(b_0x)\,,   \label{line3sol}
\ee
 where $C = -3$ corresponds to $n = 1$ or $w_{\gamma} = -2/3$, $E=c_0+2 b_0^2$ corresponds to $D_1 = {-12(c_0+2b_0^2)}/{\kappa^2}$   and $F = 2b_0^2/e_0$ corresponds to $k = {-4b_0^2}/{e_0}$. This condition demonstrates major fluid with $w_{\gamma} = -2/3$. This leads us to
 \be
 x(t)= \f{1}{b_0}{\arcsinh(e^{e_0b_0(t-t_0)})}\,, \;\;\;\;\;\text{and}\;\;\;\;\;
  u(t) = \frac{e_0e^{e_0b_0(t-t_0)}}{\sqrt{1+e^{2e_0b_0(t-t_0)}}}.
  \ee
where $b_0 x > 0$.
 The scale factor is hence
  \be
  a(t) = \f{1}{e_0^2}  \l[1+e^{-2e_0b_0(t-t_0)}\r] ,
  \ee
where $e_0 \neq 0$.
The redshift can be determined as
\be
z(t) = \frac{2}{e^{-2e_0b_0(t-t_0)}+1} - 1\,,
\ee
and there is a relation
\be
t-t_0  =  \f{-1}{2 e_0 b_0} \ln \l(\f{2}{z+1} -1  \r)\,,
\ee
whereas $z < 1 $.
The Hubble rate as function of time and redshift are
\bea
H(t) &=&  \frac{-2e_0b_0}{1+e^{2e_0b_0(t-t_0)}}\,, \\
H(z) &=&  e_0 b_0 (z-1)\,.
\eea
The barotropic fluid of this case is non-realistic with $w_{\gamma} = -2/3$.

\subsection{Solution 4}
 The exact solution is
 \be
 u(x) = e_0e^{-x\sqrt{-c_0}}-b_0e^{x\sqrt{-c_0}}\,,    \label{sol4ux}
 \ee in this case. The constant
 $E = c_0 < 0, F=0$ and $C$ is arbitrary, hence $D_1 = -12 c_0/n^2\k^2 > 0$. The results are
 \bea
 x(t)  & = &  \f{1}{\sqrt{-c_0}} \ln \l\{ \sqrt{\f{e_0}{b_0}} \tanh\l[\sqrt{-e_0 b_0 c_0} (t - t_0)   \r]  \r\}\,,  \\
 u(t)  & = &   \f{2 \sqrt{e_0 b_0}}{\sinh\l[2 \sqrt{-e_0 b_0 c_0} (t - t_0)  \r]}\,,
 \eea
and
 \bea
 a(t)  & = &   \l\{ \f{\sinh\l[2\sqrt{-e_0 b_0 c_0} (t-t_0)  \r]}{2\sqrt{e_0 b_0}} \r\}^{2/n}\,, \\
 H(t)  & = &      \f{4}{n}  \sqrt{-e_0 b_0 c_0} \coth\l[  2\sqrt{-e_0b_0c_0} (t- t_0)  \r]\,.
 \eea
Conditions need to be satisfied are $e_0, b_0$ must have the same sign and $n \neq 0$, i.e. $w_{\gamma} \neq -1$.
Having non-zero $n$ with $F=0$ implies $k=0$ (flat geometry).
The redshift $z$ is found to be constant, i.e. $z = -1$ hence there is no time-redshift relation.


\subsection{Solution 5}
 The exact solution is
 \be
 u(x) = \f{e_0}{x} e^{c_0 x^2 /2}\,,
 \ee in this case. The constants
 $E = c_0+b_0 < 0, F=0 = -nk/2$ and $C$ is arbitrary, hence $D_1 = -12 (c_0+b_0)/n^2\k^2 > 0$. Results are
 \bea
 x(t)  & = &  \sqrt{\f{-2}{c_0} \ln\l[-e_0 c_0 (t-t_0)\r]}\,, \\
 u(t)  & = &  \f{- 1}{(t-t_0)\sqrt{-2 c_0 \ln\l[ -e_0c_0(t-t_0) \r]}}\,,  \\
a(t)   & = & \l\{(t-t_0)^2 \l(  -  2 c_0 \ln \l[-e_0 c_0 (t-t_0)  \r] \r)  \r\}^{1/n} \,,\\
H(t)   &  = &  \f{1}{n(t-t_0)} \l\{  \f{1}{\ln\l[-e_0c_0(t-t_0) \r]}  + 2         \r\}\,,
 \eea
 where $c_0 < 0, n \neq 0$. At $t = t_0$, $a$ is indeterminate therefore there is no time-redshift relation.

\subsection{Solution 6}
\begin{figure}[t]  \begin{center}
\includegraphics[width=5.0cm,height=3.3cm,angle=0]{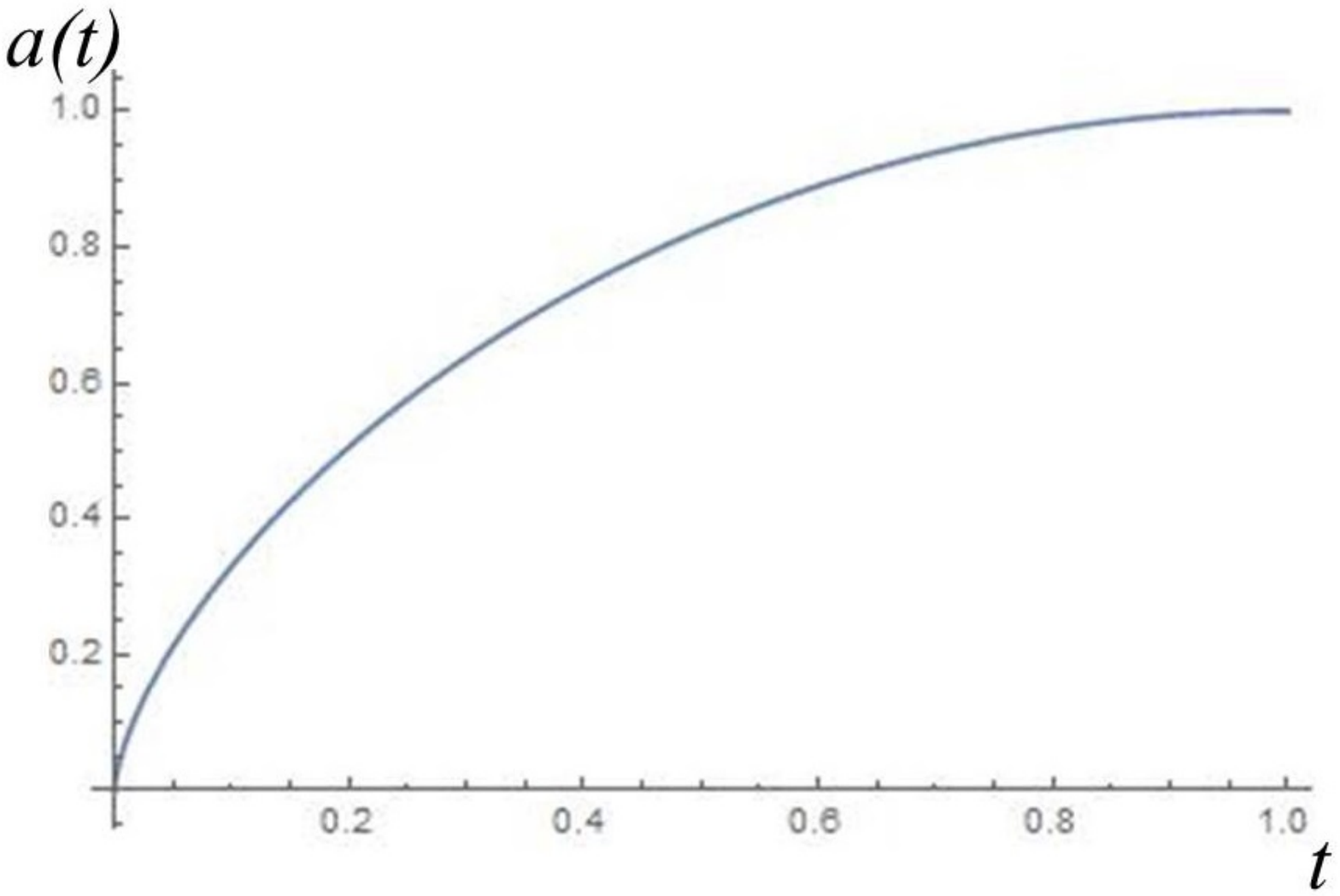}  \end{center}
\caption{Scale factor $a(t)$ of the solution 6    \label{fig1}}  \end{figure}

 The exact solution is
 \be
 u(x) = {-e_0} \cosh^2(b_0 x)\,,   \label{case6sol}
 \ee in this case. Other conditions are $E = c_0 - 2b_0^2  < 0$, $F=0$ (i.e. $k=0$), arbitrary $C$ so that
 $D_1 = -12 (c_0 - 2 b_0^2)/n^2\k^2 > 0$. The results are
 \be
 x(t)  =  \f{1}{b_0} \arctanh\l[ -e_0 b_0 (t-t_0)  \r]  \;\;\;\;\;\text{and}\;\;\;\;\;
 u(t)  =  \f{-e_0}{1 - e_0^2 b_0^2 (t-t_0)^2}\,,
 \ee
and the scale factor, redshift and Hubble rate are
\bea
a(t) & = &     \l[  \f{1 - e_0^2 b_0^2 (t-t_0)^2}{-e_0} \r]^{2/n}\,, \;\;\;\;\; a(z) \,=\, \f{1}{e_0^{2/n}(z+1)}  \\
 z(t)  & = &  \f{1}{\l[  1 - e_0^2 b_0^2 (t-t_0)^2   \r]^{2/n}}   - 1\,, \\
H(t)     &  =  &   \f{-4}{n} \l[ \f{e_0^2 b_0^2 (t-t_0)}{1 - e_0^2 b_0^2(t-t_0)^2}    \r]\,,   \\
H(z)    &  =  &   \f{-4}{n} |e_0||b_0|\sqrt{z(z+1)}\,,
\eea
 where $n \neq 0$. Taylor expansion of the solution (\ref{case6sol}) is
 \be
 u(x) =  -e_0 \l[ 1 + b_0^2(x-x_0)^2  +  \f{b_0^4}{3} (x-x_0)^4  +  \ldots  \r].   \label{case6tl}
 \ee
Compare to the power-law expansion solution $a \sim  t^q$ (with constant $q$) which corresponds to \cite{Gumjudpai:2007qq}
\be u(x)_{\text{power-law}}  =  \l[\f{(2-qn)}{2}(x-x_0)\r]^{qn/(qn - 2)}   \label{upowerlaw}
   \ee
for $n=3$ (dust), we found that the second and third terms of
the equation (\ref{case6tl}), i.e $b_0^2(x-x_0)^2$ and $[{b_0^4}(x-x_0)^4]/3$
correspond to  $ u(x)_{\text{power-law}} $ with $q= 4/3$  and $q= 8/9$ respectively. Density parameters are
\be
\Omega_1(z)  =  \f{- 3 D_1 \kappa^2}{16 b_0^2} \l[ \f{1}{(z+1)^{-3/2} -1}   \r],   \;\;\;\;\;\;     \Omega_2(z)  =  \f{- 3 D_2 \kappa^2}{16 b_0^2}\l\{ \l[ \f{  e_0^{2/3}  (z+1) }{(z+1)^{-3/2} -1}  \r]  \r\}
\ee
where $ \Omega_{\phi}(z) = 1  - \Omega_1 - \Omega_2$. Plot of $a(t)$ and $\Omega_{\phi}(z)$ are in figures \ref{fig1} and \label{fig2}. They do not resemble current observation
which suggests acceleration and present value of scalar field density parameter, $\Omega_{\phi,0}\sim 0.7$.

\begin{figure}[t]  \begin{center}
\includegraphics[width=5.0cm,height=3.3cm,angle=0]{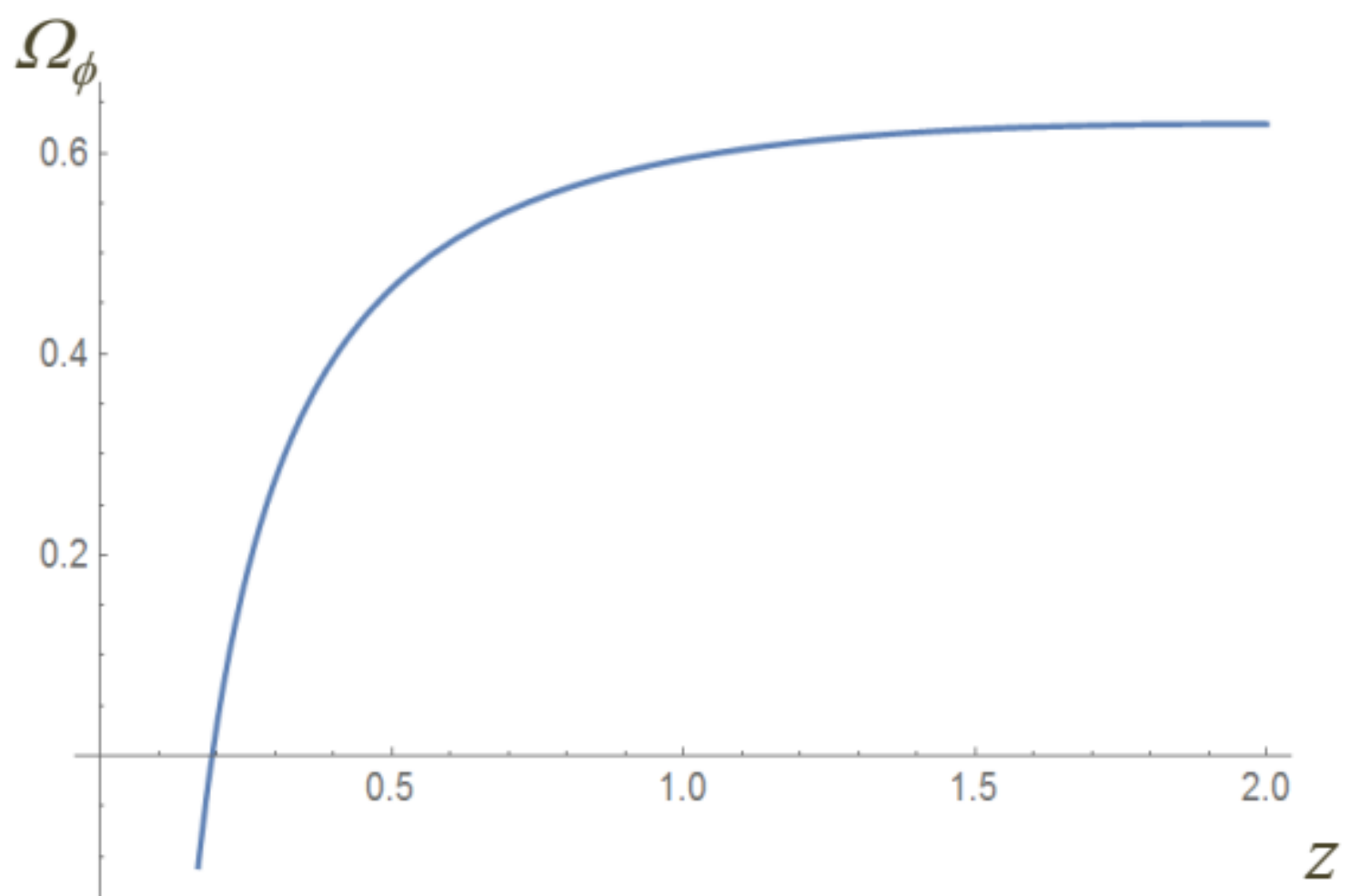}  \end{center}
\caption{Scalar field density parameter $\Omega_{\phi}(z)$ of the solution 6 plotted versus redshift.   \label{fig2}}  \end{figure}

\subsection{Solution 7}
 The exact solution is
 \be
 u(x) = \f{e_0}{x^{b_0}}\,.
 \ee Other conditions are $E = c_0$, $F=0$, arbitrary $C$. We need $c_0<0$ such that
 $D_1 = -12 c_0/n^2\k^2 > 0$. The results are
 \be
 x(t)  =  \l[e_0(b_0+1)(t-t_0)\r]^{1/(b_0+1)}  \;\;\;\;\;\text{and}\;\;\;\;\;
 u(t)  =  e_0 \l[ e_0(b_0+1)(t-t_0)   \r]^{-b_0/(b_0+1)}\,,
 \ee
and the scale factor, redshift and Hubble rate are
\bea
a(t) & = &   \f{1}{e_0^{2/n}}  \l[ e_0 (b_0+1)(t-t_0) \r]^{2b_0/[n(b_0+1)]} \,, \\
H(t)     &  =  &   \f{2 b_0}{n(b_0+1)(t-t_0)}\,,
\eea
 where $n \neq 0$. As $a(t_0) = 0 $, $z(t) = -1$ hence there is no time-redshift relation.

\subsection{Solution 8}
 Apart from the solution given by J.~D'Ambroise \cite{DAmbroise:2010dgl}, we tried solutions in form of $\cosh (b_0 x)$ and
 $\sinh (b_0 x)$ and found that they are not solutions. However we found that
 \be
 u(x) = {-e_0} \sinh^2(b_0 x)\,,    \label{case8sol}
 \ee is also a solution with
 \be
 P(x)   =  2 b_0^2 \coth^2(b_0 x) + c_0
 \ee
 with  $E = c_0 - 2b_0^2  < 0$, $F=0$, arbitrary $C$ such that
 $D_1 = -12 (c_0 - 2 b_0^2)/n^2\k^2 > 0$. Taylor expansion of the solution (\ref{case8sol}) is
\be
 u(x) =  -e_0 \l[b_0^2(x-x_0)^2  +  \f{b_0^4}{3} (x-x_0)^4  +  \ldots  \r]   \label{case8tl}
\ee
 When comparing to the solution in the power-law expansion case, (\ref{upowerlaw}), for $n=3$ (dust), we found that the first and the second terms of
the equation (\ref{case8tl}), i.e $b_0^2(x-x_0)^2$ and $[{b_0^4}(x-x_0)^4]/3$
correspond to $q= 4/3$  and $q= 8/9$ as well. The other results are
 \be
 x(t)  =  \f{1}{b_0} \arccoth\l[e_0 b_0 (t-t_0)  \r]  \;\;\;\;\;\text{and}\;\;\;\;\;
 u(t)  =  \f{e_0}{1 - e_0^2 b_0^2 (t-t_0)^2}\,,
 \ee
and the scale factor $a(t)$, redshift $z(t)$, and Hubble rate $H(t),
H(z)$ are the same as of solution 6 so as density parameters and all other relations.

\section{Conclusions and Comments}
In this work, we express NLS formulation of FRW cosmology with canonical scalar field evolving under unspecified potential and two barotropic fluids. The first barotropic density ($D_1$) is related to NLS total energy ($E$) (see equation (\ref{ua}))
 and the second barotropic fluid density ($D_2$) contributes to additional term in $P(x)$  (see equation (\ref{PxEq})).
 The choice of not adding $D_2$ term into definition of $E$ is because $E$ must be constant in deriving solutions.
 We give a lists of Friedmann formulation variables expressed in terms of NLS variables for two barotropic fluids case. The second part of this work is to explore
 seven solutions given in \cite{DAmbroise:2010dgl}. The solutions considered in this work base on top-down deducing derivation from the equation of motion (NLS equation).
 These are solutions of the system of scalar field with barotropic fluids under NLS potential ($P(x)$) listed in table \ref{tab:7solutions}. In addition, we found one new solution which gives the same result as of the sixth solution of \cite{DAmbroise:2010dgl}.

It is noticed that previous works (\cite{Gumjudpai:2007bx}, \cite{Gumjudpai:2007qq}, \cite{Phetnora:2008mf},
\cite{Gumjudpai:2009ws}, \cite{Gumjudpai:2008mg}) assumed forms of the expansion
functions, $a(t)$. These are power-law ($a \sim  t^q$), de-Sitter ($a \sim \exp(t/\tau)$
and super-acceleration ($a \sim (t_{a} - t)^q$) (with constant $q$ and $\tau$).
These expansion functions are converted to the explicit form of NLS solutions, $u(x)$.
Although it is true that $u(x)$ are exact solutions but assuming the expansion forms
is to force the problem to take the assumed answers in a bottom-up direction of reasoning.
These alter the form of scalar potential  $V = V(u, u') = V(a, \dot{a})$ to adjust so that
 the dynamics can accommodate the assumed expansions. Hence it is not a natural procedure. This is unlike conventional derivation of which at beginning step, $V(\phi)$ is taken
 from high energy physics motivation and as a result, solutions and $\Omega_{\phi}$ are derived.

All solutions-the NLS wave functions $u(x)$ found here are non-normalizable
(a specific case of power-law expansion \cite{Gumjudpai:2007qq} was also claimed to correspond to non-normalizable NLS wave function). Hence it can not be probabilistically interpreted. The NLS total energy $E$ is negative. The time-independent NLS formulation interpretation in quantum cosmology that $u(x)$ and $E$ could be the wave function and total energy
of the universe should be upgraded to the time-dependent case as in the NLS formulation reported in \cite{Lidsey:2013osa}. It is hopeful that describing Friedmann cosmology with time-dependent NLS formulation would give deeper physical insight and more realistic solutions of the problem.    This is awaiting for further investigation.

\section*{Acknowledgments}
We thank James E. Lidsey for discussion. This work is supported by a TRF Basic Research Grant no. BRG6080003 under
TRF Advanced Scholar scheme of the Thailand Research Fund and the Royal Society-Newton Advanced Fellowship of the Newton Fund (NAF-R2-180874).
BG thanks Center for Cosmology and Particle Physics, New York University for hospitality to where partial works
were completed. Prof. Dr. Sujin Jinahyon Foundation is acknowledged for traveling grant to present this work at the 10th Aegean Summer School.


\end{document}